\documentclass[11pt]{article}
\usepackage[margin=1in]{geometry}
\usepackage{amsmath,amsfonts,amssymb}
\usepackage{algorithmic}
\usepackage{algorithm}
\usepackage{array}
\usepackage{textcomp}
\usepackage{url}
\usepackage{verbatim}
\usepackage{graphicx}
\usepackage{subfig}
\usepackage{cite}
\usepackage{booktabs}
\usepackage{makecell}
\usepackage[table]{xcolor}
\usepackage{tabularx}
\usepackage{svg}
\usepackage{hyperref}

\newcommand{\modelnamecnn}[0]{CNN}
\newcommand{\modelnameresnet}[0]{ResNet}

\newcommand{\modelnametransformershort}[0]{CNNTransf.}

\begin{document}

\title{Human Activity Recognition Based on Electrocardiogram Data Only}

\author{
Sina Montazeri$^{1}$, Waltenegus Dargie$^{2}$, Yunhe Feng$^{1}$, Kewei Sha$^{1}$\\
$^{1}$Department of Computer Science and Engineering, University of North Texas, USA\\
$^{2}$Faculty of Computer Science, Technische Universit{\"a}t Dresden, Germany\\
\texttt{SinaMontazeri@unt.edu}, \texttt{Yunhe.Feng@unt.edu}, \texttt{Kewei.Sha@unt.edu},\\
\texttt{waltenegus.dargie@tu-dresden.de}
}

\date{}

\maketitle

\begin{abstract}
Human activity recognition is critical for applications such as early intervention and health analytics. Traditional activity recognition relies on inertial measurement units (IMUs), which are resource intensive and require calibration. 
% Although electrocardiogram-based methods (ECG) have been explored, they typically serve as supplements to IMUs, as reliably correlating the cardiac load with the physiological workload remains a challenge. We propose the first ECG only methodology for activity recognition, establishing a novel paradigm for low-power, simplified wearable systems. 
Although electrocardiogram (ECG)-based methods have been explored, these have typically served as supplements to IMUs or have been limited to broad categorical classification such as fall detection or active vs. inactive in daily activities. In this paper, we advance the field by demonstrating, for the first time, robust recognition of activity only with ECG in six distinct activities, which is beyond the scope of previous work.
We design and evaluate three new deep learning models, including a CNN classifier with Squeeze-and-Excitation blocks for channel-wise feature recalibration, a ResNet classifier with dilated convolutions for multiscale temporal dependency capture, and a novel CNNTransformer hybrid combining convolutional feature extraction with attention mechanisms for long-range temporal relationship modeling. Tested on data from 54 subjects for six activities, all three models achieve over 94\% accuracy for seen subjects, while CNNTransformer hybrid reaching the best accuracy of 72\% for unseen subjects, a result that can be further improved by increasing the training population. This study demonstrates the first successful ECG-only activity classification in multiple physical activities, offering significant potential for developing next-generation wearables capable of simultaneous cardiac monitoring and activity recognition without additional motion sensors.
\end{abstract}


\section{Introduction}

    \label{sec:intro}
    The Internet of Things (IoT) paradigm revolutionizes healthcare delivery through interconnected smart devices that enable continuous, personalized patient monitoring \cite{yang2022review}. According to the World Health Organization, cardiovascular diseases are the leading cause of death worldwide and claim almost 18 million lives per year~\cite{kaptoge2019world}. One-third of these deaths occur prematurely in people under 70 years of age. 
    This stark statistic underscores the urgent need for novel IoT-based healthcare solutions \cite{pang2018introduction, sivapalan2022annet}. Among these, wireless electrocardiogram (ECG) technology integrated within IoT frameworks has emerged as a cornerstone for providing continuous cardiac surveillance, capable of adapting to individual patient needs and seamlessly integrating into their lifestyle patterns \cite{tekeste2018ultra}.
    
    Human activity recognition (HAR) is a powerful tool in the fight against cardiovascular diseases, particularly for early intervention and health analytics. Traditional activity recognition relies mainly on inertial measurement units (IMUs), which are resource-intensive and require calibration \cite{wang2013learning, Wang2020WearableSH, AlKharji2023IMUbasedHA, Zhang2024LSTM}. 
    
    Although electrocardiogram-based methods (ECG) have been explored, they either identify limited activities or generally serve as complements to IMU, as the reliable correlating of cardiac load with physiological workload remains a challenge \cite{demrozi2020human,Melillo2015WearableTA,Bhoraniya2014MachineIB,ren2024clinical,ahmad2022classification,rajesh2021human}. In fact, most contemporary research combines ECG with additional sensor modalities to improve performance. Ren \textit{et al.}~\cite{ren2024clinical} demonstrated that ECG significantly improves accelerometer-based approaches, while Ahmad \textit{et al.} and Rajesh \textit{et al.}\cite{ren2024clinical, rajesh2021human} used a multisensor fusion with ECG, photoplethysmography, and IMUs.
        
    Melillo \textit{et al.}\cite{Melillo2015WearableTA} successfully applied deep transfer learning to ECG-based fall detection with 98.44\% accuracy; however, their approach remained limited to distinguishing falls from general daily activities rather than recognizing specific physical activities. Bhoraniya \textit{et al.} \cite{Bhoraniya2014MachineIB} classified body movements from motion artifacts in ambulatory ECG with 89.07\% accuracy, but treated these movements as artifacts to be filtered rather than meaningful activities to be recognized. These studies consistently position the ECG as a complementary signal rather than investigating its potential as the primary modality for comprehensive activity recognition. 
    
    This research gap motivates our investigation into ECG-only human activity recognition across six distinct physical activities. It establishes the first systematic evaluation of cardiac signals as the sole input for comprehensive activity classification without the need for additional motion sensors. Our research supports the development of next-generation IoT healthcare devices with fewer sensors to reduce costs and calibration, lower power consumption, and improved patient acceptance while maintaining the comprehensive physiological evaluation capabilities essential for personalized healthcare delivery. The contributions of this paper can be summarized as follows:
    
    \begin{itemize}
        \item We present the first comprehensive methodology for human activity recognition using electrocardiogram data as the sole input modality that classifies six distinct physical activities (sitting, standing, walking, skipping, running, climbing stairs) without requiring additional motion sensors. We demonstrate the feasibility of simplified wearable health monitoring systems that can provide simultaneous cardiac monitoring and activity recognition through a single sensor modality, with significant implications for reduced device complexity, power consumption, and improved clinical interpretation of cardiac data.
        
        \item We develop and systematically validate three deep learning models specifically adapted for ECG-based activity classification, including a CNN with Squeeze-and-Excitation blocks, a ResNet with dilated convolutions, and a novel CNNTransformer hybrid that combines convolutional feature extraction with attention mechanisms for temporal relationship modeling.
        
        \item We conduct a comprehensive performance evaluation using a six-activity dataset collected from 54 subjects. The results validate the effectiveness of our design, with all three deep learning models achieving over 94\% accuracy for seen subjects. Additionally, we present the first systematic investigation into the impact of training set size on ECG-based activity recognition. CNNTransformer hybrid has proven scalability and achieves 72\% accuracy in the holdout test set (containing unseen subjects), with performance shown to improve as the training population increases.
    \end{itemize}
    
    The remainder of the paper is organized as follows. In Section~\ref{sec:related} we review papers related to our work. In Section \ref{sec:methodology} we present our methodology and address data acquisition and preprocessing, as well as system modeling. In Section~\ref{sec:results} we evaluate our model and present quantitative results. Section~\ref{sec:discussion}, we discuss the overall system performance in terms of the various factors, as well as limitations and challenges. Finally, in Section \ref{sec:conclusion}, we conclude the paper and outline future work.

    \section{Related Work and Motivation}
    \label{sec:related}
    
    The application of electrocardiogram has evolved significantly beyond traditional cardiac diagnostics. Recent research has explored diverse applications including activity recognition, biometric identification, stress detection, and sleep analysis. This section reviews the current literature in ECG-based human activity recognition and related applications.

    \subsection{Activity Recognition} 
    
    The electrocardiogram has been shown to be useful in detecting various physiological and psychological states beyond physical activity. Tanwar \textit{et al.}\cite{Tanwar2024WearablesBP} proposed a deep learning model for ECG-based stress detection with a hybrid Long Short-Term Memory (LSTM) and gated recurrent unit neural network. Sleep analysis represents another significant application area. Feng \textit{et al.} \cite{feng2021sleep} demonstrated that single channel overnight ECG data could serve as an effective screening tool for sleep-disordered breathing by using single-lead ECG data to detect sleep apnea.
    
    However, few studies have attempted to detect human activities using ECG alone. Melillo \textit{et al.} \cite{Melillo2015WearableTA} investigated ECG-based fall detection and showed that certain physiological events such as falls can be distinguished from other physical activities based on ECG patterns; their model achieved 98. 44\% precision in classifying falls, daily activities and non-activities. Their methodology used deep transfer learning with pre-trained CNNs (AlexNet and GoogleNet) applied to continuous wavelet transform scalograms generated from filtered ECG signals. Bhoraniya \textit {et al.}\cite{Bhoraniya2014MachineIB}, on the other hand, focused on classifying body movements based on motion artifacts in ambulatory ECG. Their model achieved an accuracy of 89.07\% in discriminating activities such as hand movements, hip rotations, and the transition from sitting to standing. Their approach used adaptive filtering to extract motion artifacts, followed by the extraction of Gabor transform characteristics \cite{qian2002discrete} and the classification of multilayer perceptrons.
    
    A significant portion of current research combines ECGs with other sensor modalities to improve activity recognition. In \cite{Mahmud2020HumanAR}, the authors investigated human activity recognition using a wearable patch that combines a triaxial accelerometer and ECG sensors. They showed that additional use of an ECG significantly improves results compared to accelerometer-only approaches. Similarly, Ahmad \textit{et al.} \cite{ahmad2022classification} used deep residual networks to classify physical exercise activities using ECG, photoplethysmography (PPG), and IMU. In \cite{rajesh2021human}, the authors conducted a comparative study to evaluate the individual and combined contributions of accelerometer, ECG, and PPG signals to HAR and discuss the relative importance of different sensor modalities.

    \subsection{Research Gaps and Contributions}
        The literature reveals several important gaps that our work addresses. Most existing ECG-based activity recognition systems rely heavily on multi-sensor platforms. Studies using ECGs for HAR typically use them as a complementary signal to traditional motion sensors, rather than investigating ECGs as the primary modality. For example, Bhoraniya \textit{et al.}\cite{Bhoraniya2014MachineIB} treated body movements as artifacts to be classified rather than as meaningful activities to be recognized. Our research addresses this gap by providing the first comprehensive investigation of human activity recognition based solely on ECGs during six different physical activities. The systematic comparison of CNN, ResNet, and CNNTransformer architectures provides new information on the feasibility and limitations of cardiac signal-based activity classification. This work demonstrates that meaningful activity information can be extracted from only ECG signals and therefore establishes a new paradigm for simplified wearable cardiac monitoring systems. Reducing the number of sensors reduces costs, system overhead, and calibration requirements and improves patient acceptance. 

\section{Methodology}
\label{sec:methodology}
   
   \begin{figure*}[htbp]
        \centering
        \includegraphics[width=0.85\textwidth]{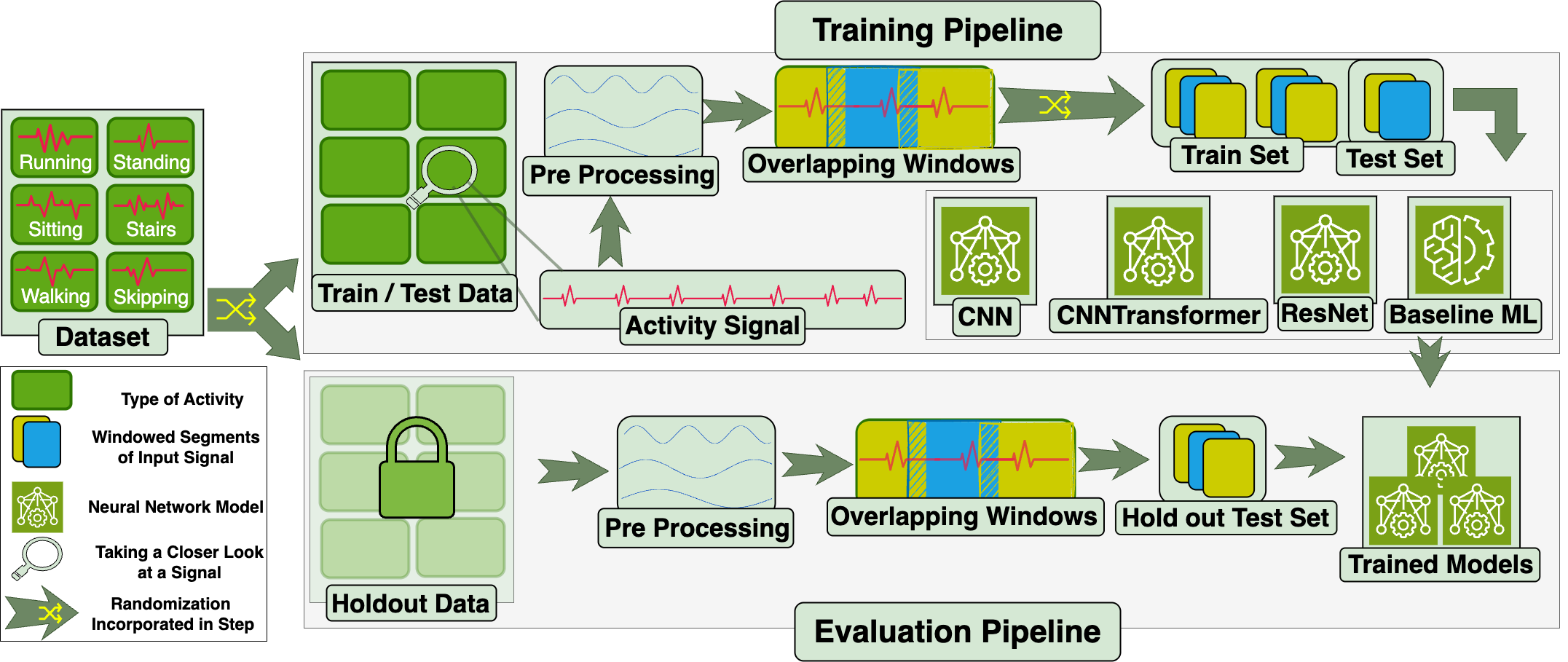}
        \caption{The ECG-only human activity recognition framework.}\label{fig:overview}
    \end{figure*}

    Our methodology, as demonstrated in Figure~\ref{fig:overview}, establishes a comprehensive framework for human activity recognition using electrocardiogram data as the only input modality. 
    
    Raw ECG data undergo systematic preprocessing to enhance signal quality and extract meaningful features. It also addresses motion artifacts and physiological noise while preserving activity-relevant cardiac patterns.
    
    We evaluated generalization by using subject-wise data splits, which test whether the models can recognize activities in individuals they had not seen during training. In all of our experiments, we create two randomized subsets of individuals in our dataset with 80\% for training and 20\% of individuals reserved as holdout test data.
    
    During training, we develop three distinct neural network architectures specifically adapted for ECG-based activity classification. The CNN classifier incorporates Squeeze-and-Excitation blocks for enhanced channel-wise feature recalibration. The ResNet classifier employs dilated convolutions within residual blocks to capture multiscale temporal dependencies. The CNNTransformer hybrid combines convolutional feature extraction with attention mechanisms to model both local morphological patterns and long-range temporal relationships in cardiac signals. As baseline models, we train classical machine learning models, and in the final step we evaluate each model's performance using the holdout test set with data from individuals never seen before.

    \subsection{Data Acquisition and Preprocessing}\label{sec:preprocessing}
\begin{figure}
        \centering
            \includegraphics[width=0.6\columnwidth]{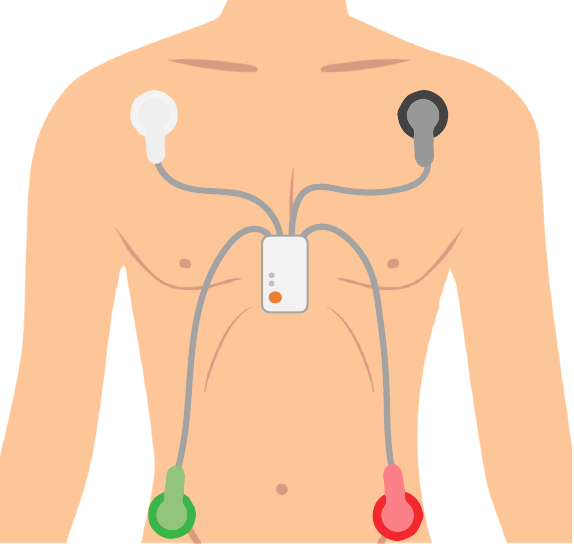}
        \caption{The placement of the Shimmer platform and its ECG electrodes.}
        \label{fig:placement}
    \end{figure}

        We used the Shimmer platform (version 3)\footnote{\url{https://shimmersensing.com/product/consensys-ecg-development-kits/}.}, a 5-lead wireless ECG, to measure the electric activities of the heart. Figure~\ref{fig:placement} illustrates the deployment of the platform. The data presented in this work were collected with the permission of the Ethics Committee of the TU Dresden (subscription no. EK271072017). Full consent from all participants had been obtained prior to the experiments. 
    
\begin{figure}[htbp]
    \centering
    \subfloat[Original ECG segment (512Hz).]{
        \includegraphics[width=\columnwidth]{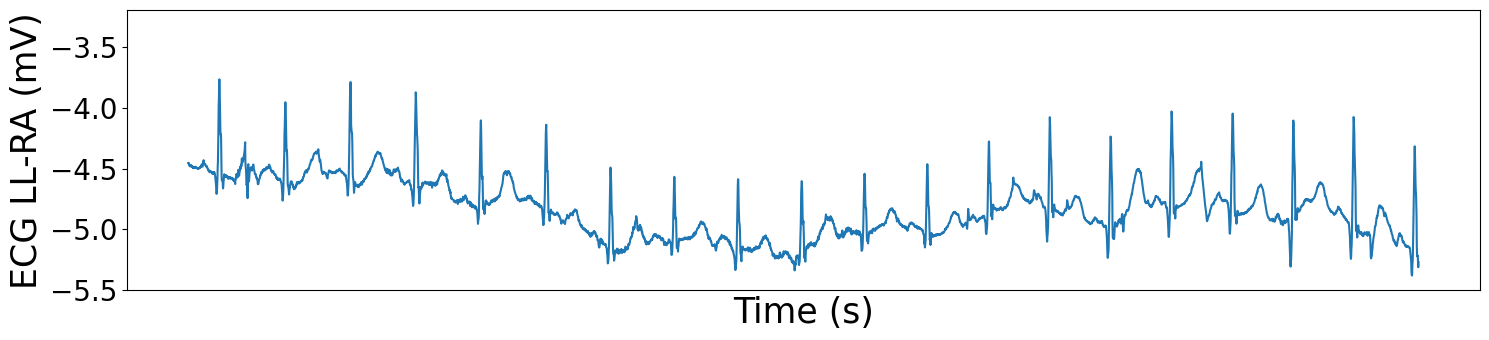}
        \label{fig:ecg_ll_la_512hz_original}
    }
    \hfill
    \subfloat[Preprocessed ECG segment (50Hz).]{
        \includegraphics[width=\columnwidth]{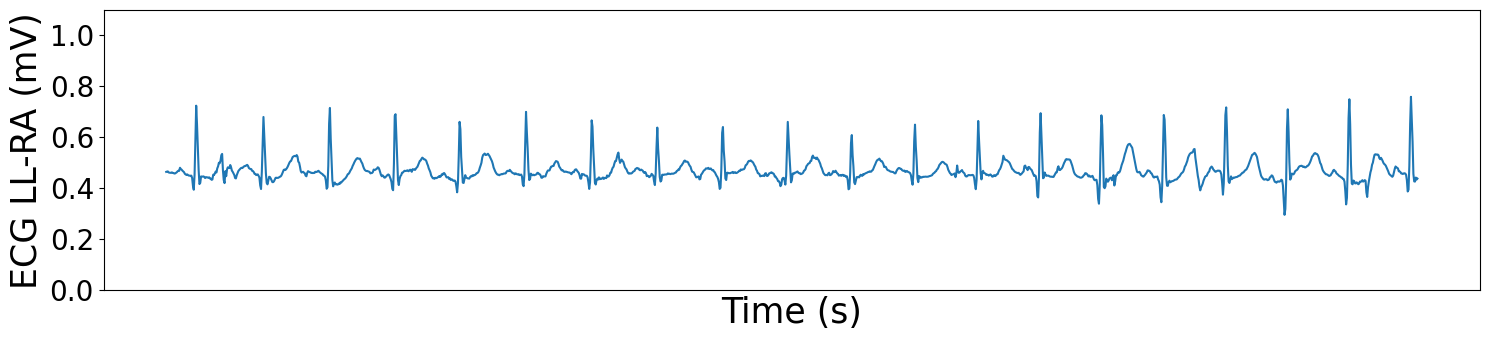}
        \label{fig:ecg_ll_la_50hz}
    }
    \caption{Comparison of original and preprocessed ECG segments during the Skipping activity.}
    \label{fig:ecg_comparison}
\end{figure}

    Two of the ECG leads provided the inputs to our models: ECG LL-LA (Lead II) and ECG LL-RA (Lead I). Figure~\ref{fig:ecg_ll_la_512hz_original} shows a five-second sample of the skipping activity originally recorded by a subject. Raw ECG data suffer from baseline drifts, resulting in low-frequency oscillations that typically occur at 0.05--1 Hz and have to be removed by applying appropriate filters. To fix this, we implemented a high-pass Butterworth filter with a cutoff frequency of 0.5 Hz and filter order of 5. We then normalize each signal and down-sampled it from 512 Hz to 50 Hz. We empirically determined that the lower resolution was sufficient for the task we aim to carry out in this paper. Figure~\ref{fig:ecg_ll_la_50hz} illustrates the sample signal after its baseline drift removal and downsampling and normalization.

\begin{figure}[htbp]
        \centering
        \includegraphics[width=\columnwidth]{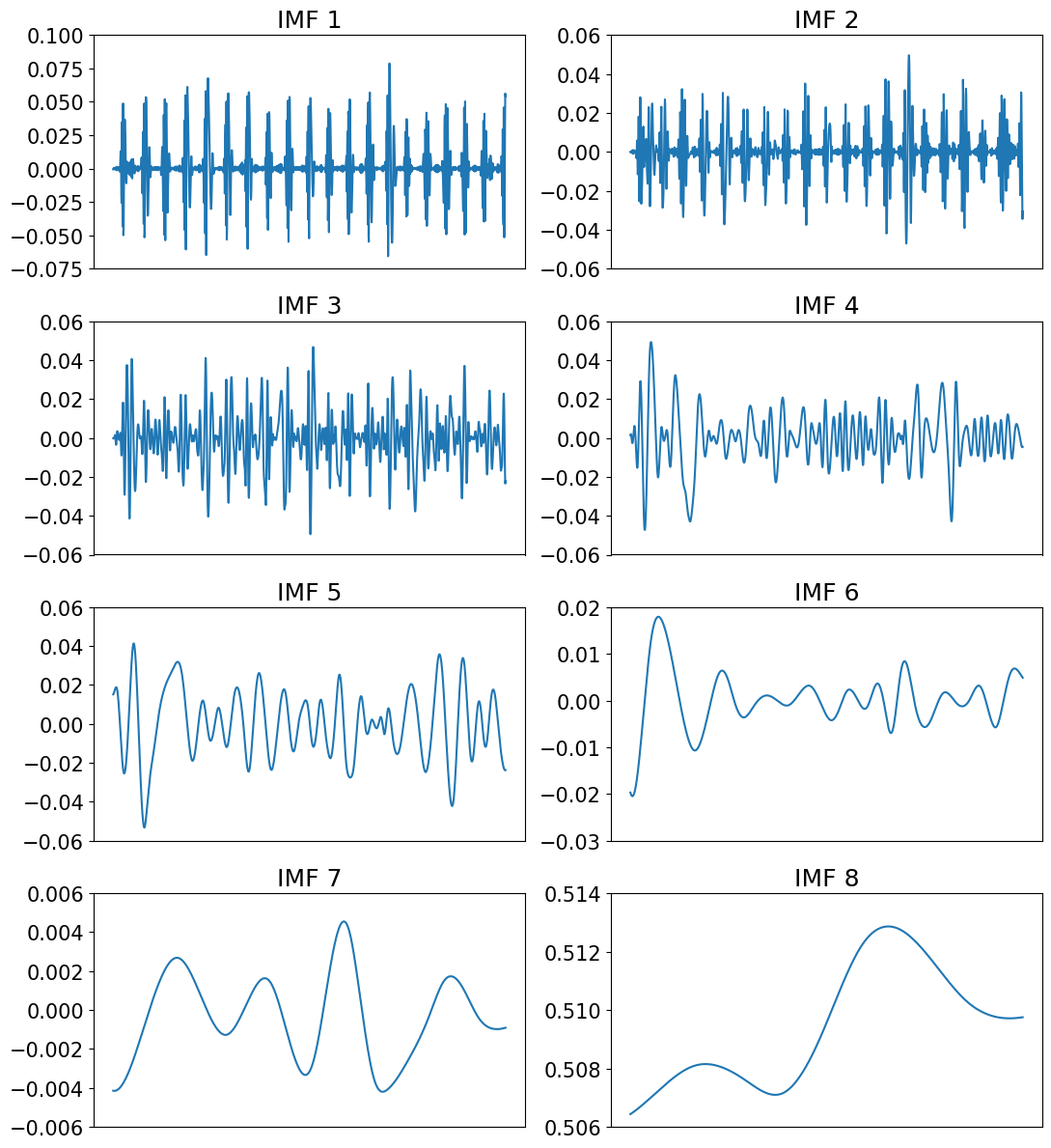}
        \caption{The 8 extracted IMFs for the example signal.}
        \label{fig:ecg_ll_la_50hz_imfs_combined}
    \end{figure}

The downsampled, denoised data are then enhanced by Empirical Mode Decomposition (EMD) \cite{Flandrin2004EmpiricalMD}. We extract Intrinsic Mode Functions (IMFs) \cite{Flandrin2004EmpiricalMD} that capture different frequency components of physiological signals. For each ECG channel (LL-LA and LL-RA), we generated eight IMFs that represent the signal at various temporal scales. Figure~\ref{fig:ecg_ll_la_50hz_imfs_combined} shows each extracted IMF for the example signal previously mentioned in Figure~\ref{fig:ecg_ll_la_50hz}. This decomposition allows the models to learn from both the original signals and their constituent frequency bands, and it provides richer feature representations for activity classification.

%    \vspace{-20pt}
%    \vspace{-20pt}

Following feature extraction, each activity data is segmented into fixed-length windows of 256 samples (this corresponds to 5 seconds at a 50 Hz sampling rate) with a sliding step of 64 samples (3.8 seconds). The overlapping windows ensure sufficient temporal context for activity recognition while maintaining adequate data coverage. The windowing process transforms continuous physiological recordings into discrete samples suitable for neural network processing.
% Figure~\ref{fig:ecg_ll_la_50hz} shows the final five-second window that is passed to our models in the training or inference pipelines.

\subsection{Neural Network Architectures}

 Classifying human activity from ECG data requires model architectures that are able to extract both local morphological features and temporal dependencies within physiological data. This work explores multiple neural network architectures specifically adapted for ECG-based activity classification. We consider established architectures such as CNNs and ResNets, while introducing key modifications tailored to the unique characteristics of ECG time series data. Additionally, we propose a novel CNNTransformer hybrid architecture that combines the local feature extraction capabilities of convolutional networks with the long-range dependency modeling strength of attention mechanisms. Each architecture is designed to address specific challenges in ECG signal processing, i.e., a CNN network with Squeeze-and-Excitation blocks for enhanced channel-wise feature recalibration; a ResNet with dilated convolutions for multi-scale temporal pattern recognition; and a hybrid CNNTransformer model for comprehensive feature extraction across different temporal scales. Figure~\ref{fig:overview} summarizes this comprehensive design framework. The following sections further elaborate on these architectures, detailing their specific adaptations for activity classification tasks.
    
\subsubsection{CNN Classifier}

One-dimensional (1D) CNNs are well suited for processing time series data like ECG segments \cite{martis2015convolutional}. They use convolutional filters to learn local patterns (e.g., waveform shapes, morphological motifs) and pooling layers to build hierarchical representations \cite{wang2015convolutional}. CNNs have demonstrated excellent performance in ECG arrhythmia classification and HAR using other sensor modalities \cite{martis2015convolutional,wang2015convolutional}.
    
    \begin{figure}[htbp]
            \centering
            \includegraphics[width=0.9\columnwidth]{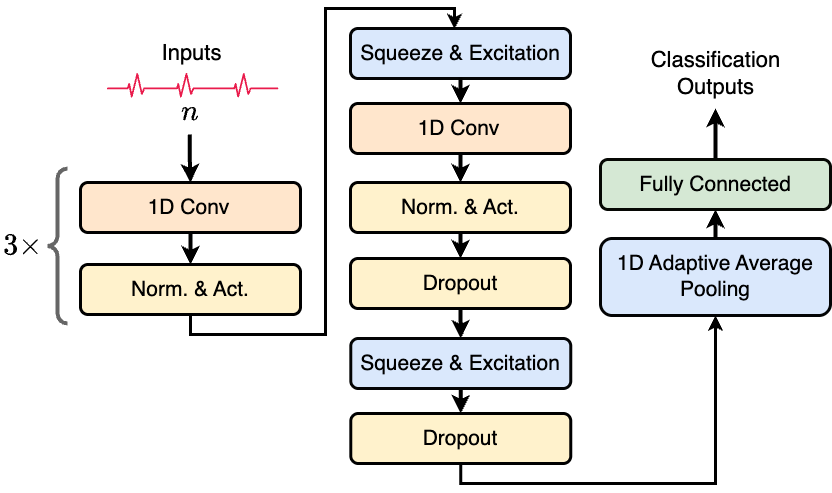}
            \caption{Architecture of the proposed CNN Classifier. The model uses convolutional blocks with increasing filter sizes and incorporates Squeeze-and-Excitation blocks for channel-wise feature recalibration, followed by global average pooling and fully connected layers for 6-class activity classification.}\label{fig:cnn-model}
        \end{figure}
    
 In order to progressively extract hierarchical features from the input time series, our CNN architecture (see Figure~\ref{fig:cnn-model}) uses a series of convolutional blocks with increasing filter sizes (64, 128, 256, 512) \cite{simonyan2015very}. After initial feature extraction by the deeper convolutional layers, to emphasize informative features and interdependencies between channels, our CNN model incorporates Squeeze-and-Excitation (SE) blocks. The SE blocks are included to adaptively recalibrate channel-wise feature responses by explicitly modeling interdependencies between channels \cite{hu2018squeeze}. The Gaussian Error Linear Unit (GELU) non-linear activation has demonstrated superior performance compared to traditional activation functions in various deep learning applications \cite{devlin2018bert}. Our CNN model uses this activation throughout its layers. This model concludes with global average pooling to condense spatial dimensions, followed by fully connected layers with dropout regularization to prevent overfitting \cite{lin2014network}.
    
\subsubsection{Residual Network (ResNet) Classifier}
\begin{figure}[htbp]
        \centering
        \includegraphics[width=0.8\columnwidth]{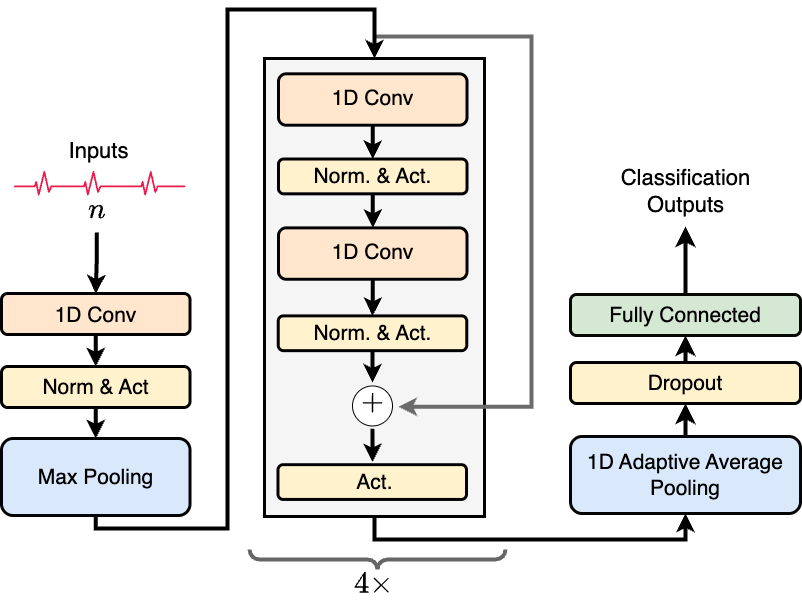}
        \caption{Architecture of the ResNet Classifier. There are 4 residual blocks incorporated. Additionally, ResNet's original classification head is replaced with global average pooling followed by a fully connected layer to support 6-class activity classification.}\label{fig:resnet-model}
    \end{figure}

Effective activity recognition from ECG data requires capturing both fine-grained waveform details and broader temporal patterns. Although convolutional neural networks (CNNs) can extract local features, increasing their depth often leads to vanishing gradients, hindering training. Residual Networks (ResNets) address this challenge by introducing shortcut connections that allow gradients to flow directly through the network, facilitating the training of deeper architectures \cite{he2016identity}. In our implementation (shown in figure~\ref{fig:resnet-model}) we adapt the ResNet architecture for one-dimensional time-series data. The model begins with an initial convolutional layer that processes the input ECG data, followed by a series of residual blocks. Each residual block comprises two convolutional layers with batch normalization and ReLU activation. A key feature of these blocks is the identity shortcut connection, which adds the input of the block to its output and  enabled the network to learn residual functions and mitigate the degradation problem associated with deep networks.

    \begin{table}[htbp]
        \centering
        \caption{ResNet Architecture Configuration}
        \label{tab:resnet-params}
        \begin{tabularx}{\columnwidth}{Xc}
        \toprule
        \textbf{Architectural Component} & \textbf{Configuration} \\
        \midrule
        Input Processing & 7×7 Conv (256 filters, stride=2) → 3×3 MaxPool \\
        Residual Layers & 4 layers: [256, 512, 1024, 2048] filters \\
        Blocks per Layer & 4 residual blocks each \\
        Multi-scale Features & Dilation rates: [1, 2, 4, 8] \\
        Regularization Strategy & Dropout: 0.4 (blocks), 0.4 (classifier) \\
        Classification Head & Global Average Pool → FC (6 classes) \\
        \bottomrule
        \end{tabularx}
    \end{table}

To capture features at multiple temporal scales, we incorporate dilated convolutions within the residual blocks. Dilated convolutions expand the receptive field without increasing the number of parameters, allowing the model to learn dependencies over longer time intervals. This is particularly beneficial for ECG data, where relevant patterns can span various time scales. After residual blocks, the model applies global average pooling to condense the feature maps, followed by a fully connected layer that produces the final activity classification. We aim to effectively balance depth and computational efficiency for processing time-series ECG data. Table~\ref{tab:resnet-params} summarizes the key parameter sizes for the ResNet Classifier model.
    
\subsubsection{CNNTransformer Hybrid Classifier}
    We previously stated that recognition activities also require effective capture of both localized waveform characteristics and longer-term temporal dependencies. Traditional convolutional approaches excel at identifying short-term morphological patterns but may overlook complex temporal relationships. Conversely, attention-based architectures, such as Transformers, are inherently adept at modeling dependencies across extended sequences, but often lack the capacity to explicitly extract local features from raw data.
        
\begin{figure}[htbp]
        \centering
        \includegraphics[width=1\columnwidth]{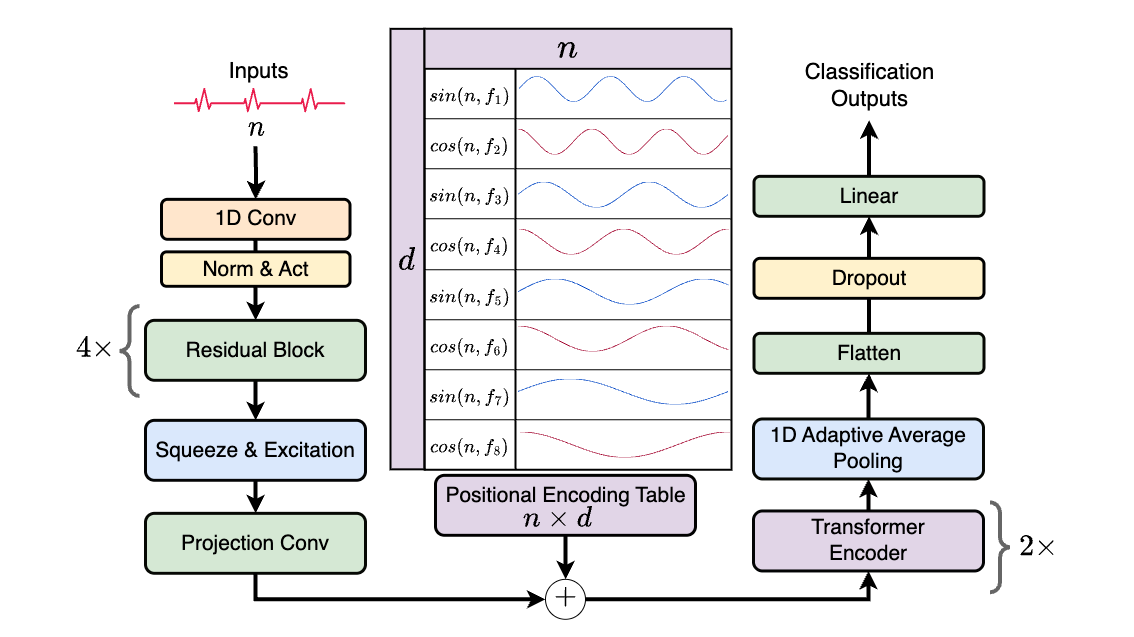}
        \caption{Architecture of the CNNTransformer classifier for an input size of length $n$ and an embedding dimension of size $d$. The model combines convolutional feature extraction with Squeeze-and-Excitation blocks, followed by Transformer encoder layers with positional encoding for temporal relationship modeling.}\label{fig:transformer-model}
    \end{figure}

    To take advantage of the complementary strengths of these architectures, we propose a hybrid CNNTransformer model, as shown in Figure~\ref{fig:transformer-model}. Initially, convolutional layers extract relevant local features from ECG waveforms, emphasizing essential morphological details. Subsequently, a Squeeze-and-Excitation (SE) block, originally introduced by Hu \textit{et al.}~\cite{hu2018squeeze}, adaptively recalibrates channel responses based on their relative importance, improving informative features. The extracted convolutional features are then projected into a high-dimensional embedding space suitable for sequence modeling. To incorporate explicit temporal structure within each discrete window, we employ sinusoidal positional encodings as introduced by Vaswani \textit{et al.}~\cite{vaswani2017attention}. Given a window length $T$ and embedding dimension $d_{\mathrm{model}}$, the positional encodings $\mathrm{PE}_{(pos,2i)}$ and $\mathrm{PE}_{(pos,2i+1)}$ are computed as follows:
        
    \begin{equation}
    \mathrm{PE}_{(pos,2i)} = \sin\left(\frac{pos}{10000^{2i/d_{\mathrm{model}}}}\right),
    \end{equation}
    \begin{equation}
    \mathrm{PE}_{(pos,2i+1)} = \cos\left(\frac{pos}{10000^{2i/d_{\mathrm{model}}}}\right),
    \end{equation}

    \noindent where $pos$ represents the position within the window ($pos \in [0, T)$), and $i$ denotes the embedding dimension index. This positional encoding explicitly embeds the relative positions of time points within each window, enabling the Transformer layers to recognize meaningful temporal arrangements and dependencies specific to each ECG segment. Transformer encoder layers, each consisting of multi-head self-attention and position-wise feedforward sub-layers, further process these temporally encoded embeddings. The self-attention mechanism identifies and emphasizes relevant dependencies in time steps, effectively distinguishing between transient physiological fluctuations and stable activity-induced patterns.
    
    The Transformer model's output is then temporally averaged to produce a robust fixed-length representation of the input window. A fully connected layer subsequently classifies this representation into distinct activity categories. This integrated design ensures a comprehensive extraction of both detailed local information and broader temporal contexts, resulting in improved accuracy in activity classification. Table~\ref{tab:transformer-params} summarizes the key parameter sizes for the Transformer Classifier.
        
    \begin{table}[htbp]
        \centering
        \caption{Parameter Sizes for the Transformer Classifier}
        \label{tab:transformer-params}
        \begin{tabularx}{\columnwidth}{Xc}
            \toprule
            \textbf{Component} & \textbf{Size} \\
            \midrule
            Embedding Dimension (\( d_{\text{model}} \)) & 512 \\
            Number of Attention Heads & 8 \\
            Number of Encoder Layers & 8 \\
            Feedforward Hidden Dimension & 512 \\
            Dropout Rate & 0.2 \\
            Output Classes & 6 \\
            Maximum Sequence Length & 12000 \\
            \bottomrule
        \end{tabularx}
    \end{table}    
    
    \subsubsection{Baseline Machine Learning Models}~\label{sec:baseline}
    % \begin{figure}[htbp]
    %         \centering
    %         \includegraphics[width=\columnwidth]{figures/machine_learning_comparison.png}
    %         \caption{Comparison of machine learning models.}\label{fig:machine-learning-comparison}
    % \end{figure}
    We implemented five classical machine learning models as baselines to evaluate ECG-based activity classification performance. These models provide comparative benchmarks for our deep learning approaches.
    
    \textbf{Support Vector Machines (SVM):} We use linear kernel SVMs (C=1.0) for their effectiveness in high-dimensional spaces. SVMs have shown success in ECG classification and multimodal human activity recognition tasks.
    
    \textbf{Random Forest (RF):} This ensemble method was introduced as a robust model against overfitting while handling high-dimensional feature sets.
    
    \textbf{k-Nearest Neighbors (kNN):} A non-parametric classifier using $k=5$ neighbors and applying majority voting among nearest neighbors in the feature space. This instance-based learner has proven to be effective for assessing physiological signals.
    
    \textbf{Decision Tree:} A single decision tree classifier that creates interpretable rules for activity classification, provides insights into feature-based decision boundaries, and serves as a simple baseline.
    
    \textbf{Logistic Regression:} A linear classifier with maximum 300 iterations that models class probabilities through logistic functions. This establishes performance bounds for linear decision boundaries in the transformed feature space. 
    
    % Each model undergoes 10 independent trials to ensure statistical reliability. Accuracy scores are recorded for each trial and visualized in Figure \ref{fig:machine-learning-comparison}. The best performing model

    \subsection{Training Methodology}
        Our training methodology consists of a unified approach to develop robust activity classification models from ECG data. To effectively evaluate generalization capability, we adopt a subject-wise splitting strategy in which activities are divided into seen and unseen categories based on participant data. This approach provides a more realistic evaluation scenario compared to random splitting and helps identify potential overfitting to individual physiological characteristics. This ensures that our models can generalize to new subjects rather than memorizing individual patterns.
        
        We use the Adam optimizer\cite{Kingma2014AdamAM} with weight decay for L2 regularization to prevent overfitting and implement a dynamic learning rate schedule to adjust the model learning rate based on validation loss plateaus. We use precision training that accelerates computation while maintaining numerical stability and weighted loss functions that adjust the contribution of each class based on its frequency to handle any potential class imbalance inherent in activity datasets. In addition, our training pipeline incorporates early stopping mechanisms that monitor validation accuracy to prevent overfitting when performance plateaus.
        
        We adopt a multistage training protocol\cite{bengio2009} with different hyperparameter configurations where each stage targets specific accuracy thresholds with customized learning rates and regularization parameters. This progressive approach allows fine-tuning of model performance while preventing catastrophic forgetting of learned features. The training stages use decreasing learning rates and increasing regularization as the model approaches optimal performance. The hyperparameter values follow established practices for deep learning optimization. Stage 1 uses a higher learning rate (4e-4) for rapid initial convergence, while the second stage reduces it to 1e-4 for precise parameter updates near convergence, consistent with learning rate scheduling strategies\cite{loshchilov2016sgdr}. Weight decay increases from 1e-4 to 1e-3 between stages to strengthen regularization as training progresses. Table~\ref{tab:training-parameters} details the specific hyperparameters used for classifier models.
        
        \begin{table}[htbp]
            \centering
            \caption{Training Hyperparameters for Classifier Models}
            \label{tab:training-parameters}
            \begin{tabularx}{\columnwidth}{
              >{\centering\arraybackslash}X 
              >{\centering\arraybackslash}X 
              >{\centering\arraybackslash}X 
              >{\centering\arraybackslash}X 
              >{\centering\arraybackslash}X 
              }
                \toprule
                \textbf{Stage} & \textbf{Learning Rate (LR)} & \textbf{Minimum LR} & \textbf{Epochs} & \textbf{Weight Decay} \\
                \midrule
                1 & 4e-4 & 1e-6 & 30 & 1e-4 \\
                2 & 1e-4 & 1e-8 & 30 & 1e-3 \\
                \bottomrule
            \end{tabularx}
        \end{table}
        
        Throughout the training process, we track performance metrics, including accuracy, loss, precision, recall, and F1 scores for both training and validation sets. Confusion matrices and misclassification analyses provide insight into model behavior and identify challenging activity transitions. All metrics are systematically recorded for experimental reproducibility and model comparison, ensuring a transparent evaluation of our methodology.

    \section{Experimental Results}
    \label{sec:results}

    We conducted two groups of experiments to systematically assess the performance generalization ability of the proposed models and the impact of the size of the training set on the accuracy of classification. This section presents an experimental evaluation of our ECG-based human activity recognition approach in three neural network architectures.

     In the first group of experiments, we split the data by subject to test how well the models work on new people. We set aside 20\% of subjects (10 individuals) as a holdout test set from the beginning. This group was never used in training and provided an unbiased test of how well models perform on completely new individuals. The remaining subjects were further divided into training and validation sets. After this data allocation step, 30 activities of each type were in the training set, 10 were in the holdout test set, and 7 in the test set. Each experimental configuration was repeated in 10 independent trials, with randomized individuals in train/test/holdout test sets to ensure statistical reliability and account for random initialization effects.

    \subsection{Performance of ML Baselines}

    \begin{figure}[htbp]
            \centering
            \includesvg[width=0.8\columnwidth]{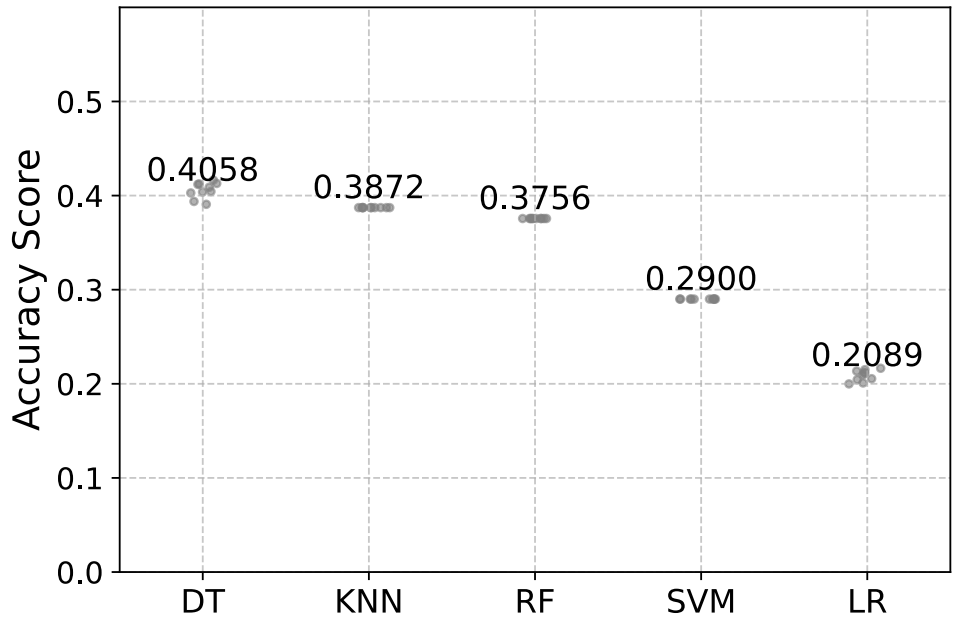}
            \caption{Performance comparison of baseline machine learning models.}\label{fig:machine-learning-comparison}
    \end{figure}

    Figure~\ref{fig:machine-learning-comparison} demonstrates the accuracy scores of the five machine learning models proposed in Section~\ref{sec:baseline}. Each model is subjected to 10 independent train/test trials to achieve statistical reliability. The results reveal that the Decision Tree achieves the highest accuracy at 40.58\%, while logistic regression performs the poorest at 20.89\%, with the remaining models clustering between 29-38\% accuracy. These traditional machine learning approaches show relatively low performance levels that fall significantly below the standards required for practical deployment, which motivated our development of more sophisticated deep learning architectures in subsequent sections.

   \subsection{Performance of CNN, ResNet, and CNNTransformer}
    After the ML baseline models, we systematically compared our three proposed architectures to identify which design principles are most effective in extracting activity-relevant patterns from ECG signals and handling intersubject physiological variability. Table~\ref{tab:model_performance_seen} shows that all three architectures achieved high performance in the test set containing seen subjects. The ResNet and CNN models both reached 96\% accuracy, with the ResNet showing slightly lower variance (±0.02) compared to the CNN (±0.03). The CNNTransformer hybrid achieved 94\% accuracy with ±0.02 variance. These results show that all architectures can effectively learn activity-specific ECG patterns when test subjects were included in the training population.

     \begin{table}[htbp]
        \centering
        \caption{Model Performance Metrics on the Test Set: Values Are Rounded Up}\label{tab:model_performance_seen}
        % \begin{tabularx}{\columnwidth}{Xcccc}
        \setlength{\tabcolsep}{5pt}
        \begin{tabularx}{\columnwidth}{ccccc}
            \toprule
            \textbf{Model} & \textbf{Accuracy} & \textbf{Precision} & \textbf{Recall} & \textbf{F1} \\
            \midrule
            \modelnametransformershort & 0.94 $\pm$ 0.02 & 0.94 $\pm$ 0.02 & 0.94 $\pm$ 0.02 & 0.94 $\pm$ 0.02 \\
            \modelnameresnet & 0.96 $\pm$ 0.02 & 0.96 $\pm$ 0.02 & 0.96 $\pm$ 0.02 & 0.96 $\pm$ 0.02 \\
            \modelnamecnn & 0.96 $\pm$ 0.03 & 0.96 $\pm$ 0.02 & 0.96 $\pm$ 0.03 & 0.96 $\pm$ 0.03 \\
            \bottomrule
        \end{tabularx}
    \end{table}
    
    The precision, recall and F1 score metrics remained consistently high in all models in the test set (seen subjects), with values ranging from 0.94 to 0.96. This consistency suggests that the models successfully learned to distinguish between the six activity classes without significant bias toward specific activities. Minimum standard deviations (0.03) in all metrics demonstrate stable performance in the 10 independent experimental trials.

    The holdout test set results presented in Table~\ref{tab:model_performance_updated} reveal a different performance landscape when models encounter completely new subjects. The CNNTransformer hybrid demonstrated superior generalization capability, achieving 72\% accuracy compared to 67\% for ResNet and 61\% for CNN. This 11 percentage point advantage over the CNN architecture and 5 percentage point lead over ResNet establishes the CNNTransformer as the most robust approach for cross-subject generalization.

    \begin{table}[htbp]
        \centering
        \caption{Model Performance Metrics on the Holdout Test Set: Values Are Rounded Up}
        \label{tab:model_performance_updated}
        % \begin{tabularx}{\columnwidth}{XXXXX}
        \setlength{\tabcolsep}{5pt}
        \begin{tabularx}{\columnwidth}{ccccc}
            \toprule
            \textbf{Model} & \textbf{Accuracy} & \textbf{Precision} & \textbf{Recall} & \textbf{F1} \\
            \midrule
            \modelnametransformershort & 0.72 $\pm$ 0.02 & 0.73 $\pm$ 0.02 & 0.72 $\pm$ 0.02 & 0.72 $\pm$ 0.02 \\
            \modelnameresnet & 0.67 $\pm$ 0.02 & 0.66 $\pm$ 0.02 & 0.67 $\pm$ 0.02 & 0.66 $\pm$ 0.02 \\
            \modelnamecnn & 0.61 $\pm$ 0.01 & 0.64 $\pm$ 0.02 & 0.61 $\pm$ 0.01 & 0.62 $\pm$ 0.01 \\
            \bottomrule
        \end{tabularx}
    \end{table}
    
    CNNTransformer had the most balanced classification performance in all types of activity. The precision and recall metrics in the CNNTransformer remained balanced at 0.73 and 0.72, respectively, and the ResNet model with 0.66 precision versus 0.67 recall. The CNN model showed the largest precision-recall gap (0.64 precision, 0.61 recall) which is not as consistent among different classes.

    The low standard deviations observed for all holdout test metrics (0.02) demonstrate that these performance differences are statistically significant and reproducible across multiple experimental trials. The CNNTransformer's consistent superiority across accuracy, precision, recall, and F1-score establishes its effectiveness for practical deployment scenarios where the system must recognize activities in previously unseen individuals.

    \subsection{Impact of Training Set Size on Model Performance}
        To understand the relationship between the diversity of the dataset and the generalizability of the model, we conducted a comprehensive investigation that examined how the size of the training set affects the performance of unseen subjects. This investigation was designed to determine whether the addition of more diverse training data improves the generalizability of the model to holdout test subjects or whether performance plateaus beyond a certain size of the dataset. 
        
        Our scaling experiments systematically varied the number of training subjects from minimal (1 subject per activity, representing 3\% of available data) to maximum (37 subjects per activity, representing 100\% of available data). The intermediate experimental points included 2, 3, 4, 7, 11, 18, and 29 subjects per activity, corresponding to 6\%, 9\%, 11\%, 19\%, 30\%, 49\%, and 79\% of the available training population, respectively. Throughout all experiments, we maintained a consistent holdout test set of 10 subjects (20\% of the total dataset) to ensure unbiased evaluation of generalization performance.

        \begin{figure}[htbp]
        \centering
        \includesvg[width=\columnwidth]{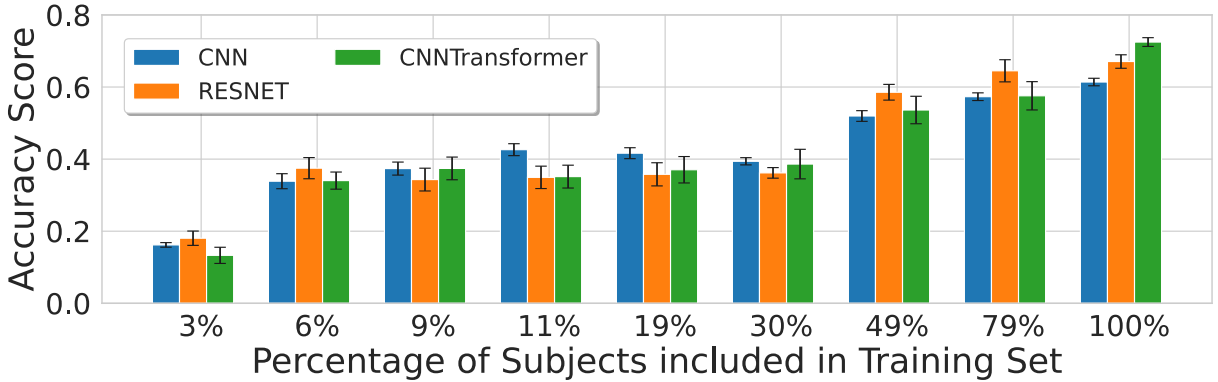}
        \caption{Model accuracy vs. number of training subjects.}
        \label{fig:performance_vs_subjects_accuracy_multimodel}
        \end{figure}

        The relationship between the model accuracy and the the number of training subjects is depicted in Figure~\ref{fig:performance_vs_subjects_accuracy_multimodel}. As shown, none of the architectures reached clear performance saturation, even when using the complete training dataset. The continued improvement in the final scaling step shows that additional subject diversity would likely yield further performance gains. This finding has important implications for future data collection efforts and indicates that ECG-based activity recognition would benefit from larger, more diverse training populations. A consistent pattern emerged across all architectures: significant performance improvements occurred around 49\% of training subjects (18 subjects per activity). This finding shows a minimum requirement to achieve clinically acceptable performance levels. Below this threshold, model uncertainty and performance variability make reliable activity recognition impractical for real-world applications.

        \begin{table}[htbp]
            \caption{CNNTransformer - Performance As Percentage of Subjects Included in Training Set Increases}
            \label{tab:single_transformer-cnn}
            \setlength{\tabcolsep}{5pt}
            \begin{tabularx}{\columnwidth}{ccccc}
                \toprule
                \centering
                \textbf{Subjects (\%)} & \textbf{Accuracy} & \textbf{Precision} & \textbf{Recall} & \textbf{F1} \\
                \midrule
                3   & 0.13 $\pm$ 0.03 & 0.14 $\pm$ 0.06 & 0.13 $\pm$ 0.03 & 0.09 $\pm$ 0.04  \\
                6   & 0.34 $\pm$ 0.03 & 0.36 $\pm$ 0.06 & 0.34 $\pm$ 0.03 & 0.30 $\pm$ 0.02  \\
                9   & 0.37 $\pm$ 0.04 & 0.36 $\pm$ 0.06 & 0.37 $\pm$ 0.04 & 0.33 $\pm$ 0.02  \\
                11  & 0.35 $\pm$ 0.04 & 0.40 $\pm$ 0.05 & 0.35 $\pm$ 0.04 & 0.33 $\pm$ 0.04  \\
                19  & 0.37 $\pm$ 0.05 & 0.36 $\pm$ 0.04 & 0.37 $\pm$ 0.05 & 0.33 $\pm$ 0.03  \\
                30  & 0.39 $\pm$ 0.05 & 0.46 $\pm$ 0.06 & 0.39 $\pm$ 0.05 & 0.38 $\pm$ 0.05  \\
                49  & 0.54 $\pm$ 0.05 & 0.59 $\pm$ 0.06 & 0.54 $\pm$ 0.05 & 0.51 $\pm$ 0.05  \\
                79  & 0.58 $\pm$ 0.05 & 0.63 $\pm$ 0.06 & 0.58 $\pm$ 0.05 & 0.57 $\pm$ 0.05  \\
                100 & 0.72 $\pm$ 0.02 & 0.73 $\pm$ 0.02 & 0.72 $\pm$ 0.02 & 0.72 $\pm$ 0.02  \\
                \bottomrule 
            \end{tabularx}
        \end{table}
        
        The CNNTransformer hybrid showed (Table~\ref{tab:single_transformer-cnn} the most dramatic performance scaling, with a 59 percentage point improvement representing the largest gain in all architectures as the size of the data increased. The learning trajectory showed different phases: low initial performance (13-37\%) through the first scaling points, a critical transition between 30-49\% of training subjects (39\% to 54\% accuracy) and the most substantial final improvement from 58\% to 72\%. Precision, recall, and F1 score metrics followed similar scaling patterns (0.14 to 0.73, 0.13 to 0.72, and 0.09 to 0.72 respectively), which shows a balanced performance across activity classes is possible only with substantial training diversity.

        Table~\ref{tab:single_resnet} shows that the ResNet model showed a 49 percentage point improvement. ResNet showed relatively stable performance during initial scaling phases, maintaining accuracy between 34\% and 37\% when using up to 30\% of training subjects. Then its performance jumps to 59\% at 49\% of training subjects and continues to improve to 67\% with the complete set. The precision, recall, and F1-score metrics scaled from 0.66, 0.67, and 0.66 respectively. Therefore the model maintained relatively balanced classification performance throughout the scaling process and robust handling of class imbalances across different training set sizes, though it was eventually not as performant as the CNNTransformer model.

        \begin{table}[htbp]
            \caption{RESNET - Performance As Percentage of Subjects Included in Training Set Increases}
            \label{tab:single_resnet}
            \setlength{\tabcolsep}{5pt}
            \begin{tabularx}{\columnwidth}{ccccc}
                \toprule
                \centering
                \textbf{Subjects (\%)} & \textbf{Accuracy} & \textbf{Precision} & \textbf{Recall} & \textbf{F1} \\
                \midrule
                3   & 0.18 $\pm$ 0.03 & 0.24 $\pm$ 0.06 & 0.18 $\pm$ 0.03 & 0.17 $\pm$ 0.04  \\
                6   & 0.37 $\pm$ 0.04 & 0.42 $\pm$ 0.03 & 0.37 $\pm$ 0.04 & 0.39 $\pm$ 0.04  \\
                9   & 0.34 $\pm$ 0.04 & 0.44 $\pm$ 0.03 & 0.34 $\pm$ 0.04 & 0.37 $\pm$ 0.04  \\
                11  & 0.35 $\pm$ 0.04 & 0.48 $\pm$ 0.05 & 0.35 $\pm$ 0.04 & 0.37 $\pm$ 0.04  \\
                19  & 0.36 $\pm$ 0.04 & 0.48 $\pm$ 0.05 & 0.36 $\pm$ 0.04 & 0.36 $\pm$ 0.05  \\
                30  & 0.36 $\pm$ 0.02 & 0.49 $\pm$ 0.03 & 0.36 $\pm$ 0.02 & 0.37 $\pm$ 0.02  \\
                49  & 0.59 $\pm$ 0.03 & 0.61 $\pm$ 0.04 & 0.59 $\pm$ 0.03 & 0.58 $\pm$ 0.03  \\
                79  & 0.65 $\pm$ 0.04 & 0.65 $\pm$ 0.04 & 0.65 $\pm$ 0.04 & 0.64 $\pm$ 0.04  \\
                100 & 0.67 $\pm$ 0.02 & 0.66 $\pm$ 0.02 & 0.67 $\pm$ 0.02 & 0.66 $\pm$ 0.02  \\
                \bottomrule 
            \end{tabularx}
        \end{table}

        The CNN architecture displayed the most variable scaling behavior among the three models. Table~\ref{tab:single_cnn} shows that CNN accuracy increased from 16\% to 61\%, a 45 percentage point improvement. However, the scaling trajectory had significant fluctuations, particularly in the middle range of training set sizes. This model's performance showed a rapid initial improvement from 16\% to 43\%, but unexpectiously declined to 39\% at 30\% of the subjects before recovering and achieving its highest gains in the final scaling phases. The accuracy jumped from 39\% to 52\% in 49\% of the subjects, then continued to improve to 61\% with the complete dataset. The improvement in F1 score from 0.12 to 0.62 was the smallest among all architectures, suggesting that the CNN model struggled the most to maintain balanced performance across activity classes as the diversity of training increased.

        \begin{table}[htbp]
        \caption{CNN - Performance As Percentage of Subjects Included in Training Set Increases}
        \label{tab:single_cnn}
        \setlength{\tabcolsep}{5pt}
            \begin{tabularx}{\columnwidth}{ccccc}
                \toprule
                \centering
                \textbf{Subjects (\%)} & \textbf{Accuracy} & \textbf{Precision} & \textbf{Recall} & \textbf{F1} \\
                \midrule
                \small
                3   & 0.16 $\pm$ 0.01 & 0.13 $\pm$ 0.01 & 0.16 $\pm$ 0.01 & 0.12 $\pm$ 0.01  \\
                6   & 0.34 $\pm$ 0.03 & 0.42 $\pm$ 0.03 & 0.34 $\pm$ 0.03 & 0.37 $\pm$ 0.03  \\
                9   & 0.37 $\pm$ 0.02 & 0.45 $\pm$ 0.02 & 0.37 $\pm$ 0.02 & 0.40 $\pm$ 0.02  \\
                11  & 0.43 $\pm$ 0.02 & 0.54 $\pm$ 0.02 & 0.43 $\pm$ 0.02 & 0.44 $\pm$ 0.02  \\
                19  & 0.42 $\pm$ 0.02 & 0.53 $\pm$ 0.01 & 0.42 $\pm$ 0.02 & 0.43 $\pm$ 0.02  \\
                30  & 0.39 $\pm$ 0.01 & 0.50 $\pm$ 0.01 & 0.39 $\pm$ 0.01 & 0.41 $\pm$ 0.01  \\
                49  & 0.52 $\pm$ 0.02 & 0.58 $\pm$ 0.01 & 0.52 $\pm$ 0.02 & 0.53 $\pm$ 0.02  \\
                79  & 0.57 $\pm$ 0.01 & 0.61 $\pm$ 0.01 & 0.57 $\pm$ 0.01 & 0.58 $\pm$ 0.01  \\
                100 & 0.61 $\pm$ 0.01 & 0.64 $\pm$ 0.02 & 0.61 $\pm$ 0.01 & 0.62 $\pm$ 0.01  \\
                \bottomrule
            \end{tabularx}
        \end{table}
        
\begin{figure}[htbp]
        \centering
        \includesvg[width=\linewidth]{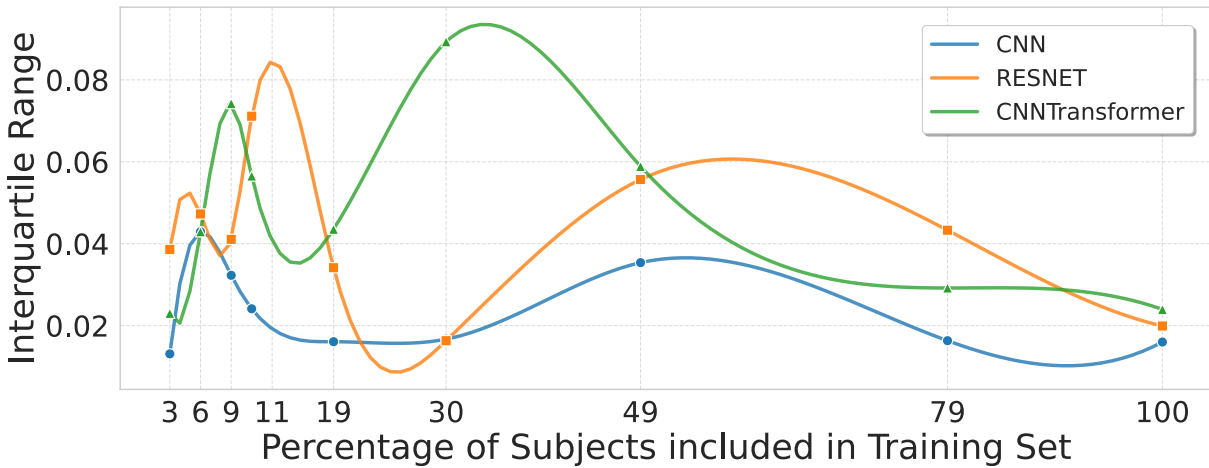}
        \caption{Comparison of model uncertainty as the number of subjects increases.}\label{fig:uncertainty_iqr_trend}
        \end{figure}

        In this experiment, we also measured the interqualtile range as a measure of uncertainty (Figure \ref{fig:uncertainty_iqr_trend}) in 10 trials. This figure shows that all architectures struggled with small training populations. The CNNTransformer model had the most dramatic fluctuations and peaked around 30\% of the training subjects before dramatically decreasing. ResNet showed a high initial uncertainty that decreased with notable spikes in subjects 69\% and 49\%. CNN maintained the most stable uncertainty profile throughout training. When using complete training sets, all models achieved low uncertainty levels, with CNN performing best, followed by ResNet and CNNTransformer. This uncertainty reduction matched accuracy improvements and confirms that larger and more diverse datasets enhance both performance and prediction reliability.

\section{Discussion}
\label{sec:discussion}

\subsection{Using ECG vs Multiple Modalities for Human Activity Recognition}

Our experimental results demonstrate the feasibility of human activity recognition using electrocardiogram data alone. The CNNTransformer hybrid model achieved 72\% accuracy on completely unseen subjects, while the ResNet and CNN models reached 67\% and 61\%, respectively. These results represent the first successful demonstration of ECG-only activity classification in six distinct physical activities, establishing a new paradigm for wearable health monitoring systems.

The performance gap between test data (96\% accuracy) and holdout test data (61--72\% accuracy) demonstrates an important point about physiological variability between individuals. Each person exhibits unique cardiac responses to physical exertion, influenced by factors such as fitness level, age, cardiac health, and individual physiology. This variability requires robust models capable of generalizing beyond training populations.

Table~\ref{tab:comparison_ecg_har} compares existing ECG-based human activity recognition studies with current work. Most previous approaches either incorporate additional sensor modalities or focus on limited activity sets. Earlier work such as \textit{Bhoraniya et al.}~\cite{Bhoraniya2014MachineIB} focused on movements of a single body part rather than physiological responses to activities, while more recent studies such as \textit{Ren et al.}~\cite{ren2024clinical} and \textit{Ahmad et al.}~\cite{ahmad2022classification} use multisensor fusion to achieve high performance. Many studies lack rigorous cross-subject validation, which is essential for demonstrating real-world applicability. The present work is the first comprehensive investigation to classify six distinct physical activities using only ECG signals, with systematic cross-subject validation that demonstrates precision 72\% in unseen subjects. This establishes an important baseline for ECG-only activity recognition and demonstrates the feasibility of simplified wearable monitoring systems that can operate without additional motion sensors.

    \begin{table*}[htbp]
    \centering      
    \caption{Comparison of ECG-Based human activity recognition studies.}
    \label{tab:comparison_ecg_har}
    \rowcolors{2}{gray!15}{white}
    \resizebox{\textwidth}{!}{%
        \begin{tabular}{m{1.3cm}m{2.2cm}m{2cm}m{2.5cm}m{1.5cm}m{2.5cm}m{2cm}m{3.5cm}}
        \hline
        \textbf{Study} & \textbf{Application Focus} & \textbf{Input Signals} & \textbf{Activities/Classes} & \textbf{Dataset Size} & \textbf{Methodology} & \textbf{Cross-Subject Performance} & \textbf{Key Contribution/Limitation} \\
        \hline
        \textbf{Our Work (2025)} & \textbf{ECG-HAR} & \textbf{2-lead ECG} & \textbf{6 activities (sitting, standing, walking, skipping, running, climbing stairs)} & \textbf{54 subjects} & \textbf{CNNTransformer hybrid, ResNet, CNN with EMD preprocessing} & \textbf{72\% accuracy (unseen subjects)} & \textbf{First comprehensive ECG-only classification across 6 distinct activities; subject-wise validation ensures generalization} \\
        
        Ren et al. (2024) & Clinical HAR & Tri-axial ACC + ECG & 7 activities (healthy), 4 activities (patients) & 45 total (20 healthy + 25 pneumoconiosis) & CNN-LSTM & Not reported as cross-subject & Multi-modal approach; clinical validation but requires additional sensors \\
        
        Rajesh et al. (2021) & Comparative HAR study & 3D-ACC + ECG + PPG & Daily activities & 15 subjects (PPG-DaLiA dataset) & Random Forest with early fusion & 86.17\% F1-score (cross-subject) & Systematic comparison of modalities; shows ECG adds 3\% improvement to ACC alone \\
        
        Melillo et al. (2021) & Fall detection & 3-lead ECG (converted to scalograms) & 3 classes (fall, daily activities, no activities) & 6 subjects (augmented to 1270 samples) & Transfer learning (AlexNet, GoogLeNet) & 97.37\% accuracy & ECG-only but limited to fall detection; heavy data augmentation \\
        
        Bhoraniya et al. (2014) & Body movement classification & Single-lead ECG (motion artifacts) & 4 movement types (left arm, right arm, waist twisting, sit-stand) & 6 subjects & MLPFF neural network with PCA and statistical features (mean, median, variance, max) & 82.59\% (adaptive filtering); 87.17\% (DWT) accuracy & DWT-based motion artifact extraction outperforms adaptive filtering for BMA classification \\
        
        Ahmad et al. (2022) & Exercise classification & ECG + PPG + IMU & Exercise activities & 8 subjects & Deep ResNet & Not reported & Multi-modal deep learning approach; details limited \\
        \hline
    \end{tabular}%
    }   
\end{table*}

\subsection{Model Architecture Performance Analysis}
    
    The performance gap between architectures on the holdout test set reveals a significant difference in their ability to handle physiological variability. The CNNTransformer's 5-7 percentage point advantage over ResNet and 11 percentage, over CNN, is due to the hybrid model's ability to learn adaptive feature weighting in temporal sequences; the transformer component identifying which cardiac response patterns are most relevant for activity classification among different individuals and compensating for intersubject physiological variations; and the CNN component detecting local ECG morphological changes. % and ResNets effectively capture hierarchical patterns through dilated convolutions, the CNNTransformer's superior holdout test set performance suggests that dynamic temporal context adjustment is essential for robust activity recognition from cardiac signals, as neither purely convolutional approach can adaptively weight feature importance based on temporal relationships across diverse physiological responses.
    
    \subsection{Impact of Dataset Size on Generalization}
    The consistent performance improvements across all training set sizes confirms what looks like the obvious, namely, that the robustness of ECG-based activity recognition steadily improves with increasing dataset size. %can be made robust by increasing benefits from rich datasets suffers from a significant data hunger, probably due to the high variability between individuals in cardiac responses to physical activities. The absence of performance saturation even at maximum training size suggests that substantially larger datasets would be required to achieve optimal generalization.
    The critical mass effect observed around 49\% of the subjects demonstrates an important threshold for practical deployment. Below this point, models exhibit unstable performance. This threshold likely represents the minimum subject diversity required to capture the range of physiological responses necessary for robust cross-subject generalization.
    
    The dramatic reduction in uncertainty with larger training sets addresses a crucial clinical concern. High model uncertainty with small datasets creates unpredictable behavior that could compromise patient safety in real-world applications. The convergence to low uncertainty levels with complete training sets suggests that reliable recognition of ECG-based activity requires a comprehensive representation of physiological diversity rather than simply achieving high average accuracy.

    \subsection{Physiological Mechanisms and Clinical Applications}
    The success of ECG-based activity recognition originates from the cardiac system's predictable responses to physical exertion, where activities such as running or climbing stairs produce distinct changes in heart rate, cardiac output, and ECG morphology that our Empirical Mode Decomposition preprocessing effectively captured across multiple frequency domains. This physiological foundation enabled high classification accuracy for activities with distinct cardiac demands, while presenting greater challenges for similar activities like standing versus sitting, which validates the approach's biological basis. 
    
    These findings have significant implications for next-generation wearable health devices that currently rely on multiple sensors, as ECG-based activity recognition could simplify device designs while simultaneously providing cardiac monitoring and activity classification. The approach offers particular advantages for cardiac patients who require continuous ECG monitoring, but whose traditional systems lack the activity context, thus limiting the clinical interpretation of cardiac events. The clinical translation potential includes cardiac rehabilitation programs that could benefit from integrated monitoring of cardiac status and physical activity levels, remote patient monitoring systems that would provide comprehensive physiological evaluations, and enhanced clinical value for existing cardiac monitoring systems, although extensive validation in diverse patient populations and regulatory approval processes remain essential prerequisites for practical implementation.
    
    \subsection{Limitations and Future Research Directions}
    This study's primary limitations include a predominantly young, healthy subject population (82\% between ages 21-24) that restricts generalizability to older adults and individuals with cardiac conditions, whose cardiac responses to physical activity vary significantly with age and health status. The controlled experimental environment with discrete activities performed for fixed durations may not reflect real-world usage scenarios that involve complex transitions and varying intensities, while motion artifacts during intense physical activities continue to degrade signal quality despite preprocessing efforts. 
    
    Future research should prioritize advanced motion artifact removal techniques that can better handle the dynamic nature of physical activities. Identification of cardiac features that remain persistent between different demographic groups and activity types represents another critical research direction. These persistent features could improve model robustness without requiring additional sensor modalities.
    Long-term validation studies in real-world environments across diverse populations remain essential. These studies should evaluate performance in older adults and individuals with various health conditions. Personalized models that use transfer learning techniques offer the potential for rapid adaptation to new users with minimal calibration data. Integration of cardiac physiology domain knowledge into model architectures presents additional opportunities for improvement while maintaining the single-sensor approach that defines this work's contribution.

\section{Conclusion}
    \label{sec:conclusion}
    
    This paper successfully demonstrates, for the first time, the classification of six distinct physical activities for human activity recognition using only ECG data. Our novel CNNTransformer hybrid model achieved 72\% accuracy on unseen subjects, outperforming two other proposed deep learning architectures. These findings establish a paradigm for simplified wearable systems that provide simultaneous cardiac monitoring and human activity recognition without additional motion sensors, thus reducing device complexity, power consumption, and cost. The CNNTransformer's effective modeling of both local ECG morphological patterns and long-range temporal dependencies to address intersubject variability, alongside the demonstrated critical need for large and diverse training datasets for robust generalization, are central to this success. Although a broader validation across diverse populations and real-world conditions is an important next step, our work represents a transformative advance in wearable health monitoring, promising substantial benefits for personalized healthcare and remote patient management.

\bibliographystyle{plain}
\bibliography{reference}

\begin{thebibliography}{10}

\bibitem{ahmad2022classification}
Zulfiqar Ahmad et~al.
\newblock Classification of physical exercise activity from ecg, ppg and imu
  sensors using deep residual network.
\newblock In {\em 2022 IEEE International Conference on Signal Processing and
  Communications (SPCOM)}, pages 1--5. IEEE, 2022.

\bibitem{AlKharji2023IMUbasedHA}
Saad AlKharji, Aysha Alteneiji, and Kin Poon.
\newblock Imu-based human activity recognition using machine learning and deep
  learning models.
\newblock {\em 2023 6th International Conference on Signal Processing and
  Information Security (ICSPIS)}, pages 62--66, 2023.

\bibitem{bengio2009}
Yoshua Bengio, J\'{e}r\^{o}me Louradour, Ronan Collobert, and Jason Weston.
\newblock Curriculum learning.
\newblock In {\em Proceedings of the 26th Annual International Conference on
  Machine Learning}, ICML '09, page 41–48, New York, NY, USA, 2009.
  Association for Computing Machinery.

\bibitem{Bhoraniya2014MachineIB}
Dixit~V. Bhoraniya and Rahul Kher.
\newblock Machine intelligence based identification of body movements in
  ambulatory ecg (a-ecg).
\newblock {\em 2014 International Conference on Medical Imaging, m-Health and
  Emerging Communication Systems (MedCom)}, pages 80--85, 2014.

\bibitem{demrozi2020human}
Florenc Demrozi, Graziano Pravadelli, Azra Bihorac, and Parisa Rashidi.
\newblock Human activity recognition using inertial, physiological and
  environmental sensors: A comprehensive survey.
\newblock {\em IEEE access}, 8:210816--210836, 2020.

\bibitem{devlin2018bert}
Jacob Devlin, Ming-Wei Chang, Kenton Lee, and Kristina Toutanova.
\newblock Bert: Pre-training of deep bidirectional transformers for language
  understanding.
\newblock In {\em Advances in Neural Information Processing Systems (NeurIPS)},
  pages 4171--4186, 2018.

\bibitem{feng2021sleep}
Kaicheng Feng, Hengji Qin, Shan Wu, Weifeng Pan, and Guanzheng Liu.
\newblock A sleep apnea detection method based on unsupervised feature learning
  and single-lead electrocardiogram.
\newblock {\em IEEE Transactions on Instrumentation and Measurement}, 70:1--12,
  2021.

\bibitem{Flandrin2004EmpiricalMD}
Patrick Flandrin, Gabriel Rilling, and Paulo Gonçalves.
\newblock Empirical mode decomposition as a filter bank.
\newblock {\em IEEE Signal Processing Letters}, 11:112--114, 2004.

\bibitem{he2016identity}
Kaiming He, Xiangyu Zhang, Shaoqing Ren, and Jian Sun.
\newblock Identity mappings in deep residual networks.
\newblock In {\em Computer Vision--ECCV 2016: 14th European Conference,
  Amsterdam, The Netherlands, October 11--14, 2016, Proceedings, Part IV 14},
  pages 630--645. Springer, 2016.

\bibitem{hu2018squeeze}
Jie Hu, Li~Shen, and Gang Sun.
\newblock Squeeze-and-excitation networks.
\newblock In {\em Proceedings of the IEEE/CVF Conference on Computer Vision and
  Pattern Recognition (CVPR)}, pages 7132--7141, 2018.

\bibitem{kaptoge2019world}
Stephen Kaptoge, Lisa Pennells, Dirk De~Bacquer, Marie~Therese Cooney, Maryam
  Kavousi, Gretchen Stevens, Leanne~Margaret Riley, Stefan Savin, Taskeen Khan,
  Servet Altay, et~al.
\newblock World health organization cardiovascular disease risk charts: revised
  models to estimate risk in 21 global regions.
\newblock {\em The Lancet global health}, 7(10):e1332--e1345, 2019.

\bibitem{Kingma2014AdamAM}
Diederik~P. Kingma and Jimmy Ba.
\newblock Adam: A method for stochastic optimization.
\newblock {\em CoRR}, abs/1412.6980, 2014.

\bibitem{lin2014network}
Min Lin, Qiang Chen, and Shuicheng Yan.
\newblock Network in network.
\newblock {\em International Conference on Learning Representations (ICLR)},
  2014.

\bibitem{loshchilov2016sgdr}
Ilya Loshchilov and Frank Hutter.
\newblock Sgdr: Stochastic gradient descent with warm restarts.
\newblock In {\em International Conference on Learning Representations}, 2017.

\bibitem{Mahmud2020HumanAR}
Tanvir Mahmud, Shaimur~Salehin Akash, Shaikh~Anowarul Fattah, Weiping Zhu, and
  M.~Omair Ahmad.
\newblock Human activity recognition from multi-modal wearable sensor data
  using deep multi-stage lstm architecture based on temporal feature
  aggregation.
\newblock {\em 2020 IEEE 63rd International Midwest Symposium on Circuits and
  Systems (MWSCAS)}, pages 249--252, 2020.

\bibitem{martis2015convolutional}
Roshan~Joy Martis, U.~Rajendra Acharya, Choo~Min Lim, K.~M. Mandana, Ajoy~Kumar
  Ray, and Chandan Chakraborty.
\newblock Convolutional neural networks for patient-specific ecg
  classification.
\newblock {\em Computing in Cardiology Conference (CinC)}, pages 557--560,
  2015.

\bibitem{Melillo2015WearableTA}
Paolo Melillo, Rossana Castaldo, Giovanna Sannino, Ada Orrico, Giuseppe~De
  Pietro, and Leandro Pecchia.
\newblock Wearable technology and ecg processing for fall risk assessment,
  prevention and detection.
\newblock {\em 2015 37th Annual International Conference of the IEEE
  Engineering in Medicine and Biology Society (EMBC)}, pages 7740--7743, 2015.

\bibitem{pang2018introduction}
Zhibo Pang, Geng Yang, Ridha Khedri, and Yuan-Ting Zhang.
\newblock Introduction to the special section: convergence of automation
  technology, biomedical engineering, and health informatics toward the
  healthcare 4.0.
\newblock {\em IEEE reviews in biomedical engineering}, 11:249--259, 2018.

\bibitem{qian2002discrete}
Shie Qian and Dapang Chen.
\newblock Discrete gabor transform.
\newblock {\em IEEE transactions on signal processing}, 41(7):2429--2438, 2002.

\bibitem{rajesh2021human}
Abdu Rajesh, Subhas~C Mukhopadhyay, et~al.
\newblock Human activity recognition: A comparative study to assess the
  contribution level of accelerometer, ecg, and ppg signals.
\newblock {\em Sensors}, 21(21):7300, 2021.

\bibitem{ren2024clinical}
Yanling Ren, Minqi Liu, Ying Yang, Ling Mao, and Kai Chen.
\newblock Clinical human activity recognition based on a wearable patch of
  combined tri-axial acc and ecg sensors.
\newblock {\em Digital Health}, 10, 2024.

\bibitem{simonyan2015very}
Karen Simonyan and Andrew Zisserman.
\newblock Very deep convolutional networks for large-scale image recognition.
\newblock {\em International Conference on Learning Representations (ICLR)},
  2015.

\bibitem{sivapalan2022annet}
Gawsalyan Sivapalan, Koushik~Kumar Nundy, Soumyabrata Dev, Barry Cardiff, and
  Deepu John.
\newblock Annet: A lightweight neural network for ecg anomaly detection in iot
  edge sensors.
\newblock {\em IEEE Transactions on Biomedical Circuits and Systems},
  16(1):24--35, 2022.

\bibitem{Tanwar2024WearablesBP}
Ritu Tanwar, Pankaj~Kumar Pal, and Ghanapriya Singh.
\newblock Wearables based personalised stress recognition using signal
  processing and hybrid deep learning model.
\newblock {\em 2024 International Conference on Computer, Electronics,
  Electrical Engineering \& their Applications (IC2E3)}, pages 1--6, 2024.

\bibitem{tekeste2018ultra}
Temesghen Tekeste, Hani Saleh, Baker Mohammad, and Mohammed Ismail.
\newblock Ultra-low power qrs detection and ecg compression architecture for
  iot healthcare devices.
\newblock {\em IEEE Transactions on Circuits and Systems I: Regular Papers},
  66(2):669--679, 2018.

\bibitem{vaswani2017attention}
Ashish Vaswani, Noam Shazeer, Niki Parmar, Jakob Uszkoreit, Llion Jones,
  Aidan~N Gomez, {\L}ukasz Kaiser, and Illia Polosukhin.
\newblock Attention is all you need.
\newblock In {\em Advances in Neural Information Processing Systems},
  volume~30, pages 5998--6008. Curran Associates, Inc., 2017.

\bibitem{wang2015convolutional}
Hongyi Wang and Bhiksha Raj.
\newblock Convolutional neural networks for human activity recognition using
  multichannel time series.
\newblock In {\em Proceedings of the Twenty-Fourth International Joint
  Conference on Artificial Intelligence (IJCAI)}, pages 3927--3933, 2015.

\bibitem{Wang2020WearableSH}
Huaijun Wang, Jing Zhao, Junhuai Li, Ling Tian, Pengjia Tu, Ting Cao, Yang An,
  Kan Wang, and Shancang Li.
\newblock Wearable sensor-based human activity recognition using hybrid deep
  learning techniques.
\newblock {\em Secur. Commun. Networks}, 2020:2132138:1--2132138:12, 2020.

\bibitem{wang2013learning}
Jiang Wang, Zicheng Liu, Ying Wu, and Junsong Yuan.
\newblock Learning actionlet ensemble for 3d human action recognition.
\newblock {\em IEEE transactions on pattern analysis and machine intelligence},
  36(5):914--927, 2013.

\bibitem{yang2022review}
Yilin Yang, Haocong Wang, Ruizhe Jiang, Xiaonan Guo, Jerry Cheng, and Yingying
  Chen.
\newblock A review of iot-enabled mobile healthcare: technologies, challenges,
  and future trends.
\newblock {\em IEEE Internet of Things Journal}, 9(12):9478--9502, 2022.

\bibitem{Zhang2024LSTM}
Ningbo Zhang, Yang Song, Dongxu Fang, Zhiwei Gao, and Yajie Yan.
\newblock An improved deep convolutional lstm for human activity recognition
  using wearable sensors.
\newblock {\em IEEE Sensors Journal}, 24(2):1717--1729, 2024.

\end{thebibliography}

\end{document}